\documentclass[aps,twocolumn,showpacs,preprintnumbers,amsmath,amssymb,floatfix]{revtex4-2}

\usepackage{graphicx}
\usepackage{longtable}
\usepackage{dcolumn}
\usepackage{bm}
\usepackage{color}
\usepackage[normalem]{ulem}
\usepackage[colorlinks=true, urlcolor=blue, linkcolor=blue, citecolor=blue]{hyperref}
\usepackage[all]{hypcap}
\newcommand{\1}{\c{c}}

\newcommand{\beq}{\begin{equation}}
\newcommand{\eeq}{\end{equation}}
\newcommand{\bea}{\begin{eqnarray}}
\newcommand{\eea}{\end{eqnarray}}

\begin{document}

\title{Itinerant versus localized magnetism in spin gapped  metallic half-Heusler compounds: 
Stoner criterion and magnetic interactions}

\author{E. \c{S}a\c{s}{\i}o\u{g}lu$^{1}$}\email{ersoy.sasioglu@physik.uni-halle.de}
\author{W. Beida$^{2,3}$}
\author{S. Ghosh$^{4}$}
\author{M. Tas$^{5}$}
\author{B. Sanyal$^{6}$}
\author{S. Lounis$^{1}$}
\author{S. Bl\"{u}gel$^{3}$}
\author{I. Mertig$^{1}$}
\author{I. Galanakis$^{7}$}\email{galanakis@upatras.gr}

\affiliation{$^{1}$Institute of Physics, Martin Luther University Halle-Wittenberg, 06120 Halle (Saale), Germany\\
$^{2}$Physics Department, RWTH Aachen University, 52062, Aachen, Germany\\
$^{3}$Peter Gr\"unberg Institut, Forschungszentrum J\"ulich and JARA, 52425 J\"ulich, Germany\\
$^{4}$Department of Physics, Central University of Kashmir,Tulmulla, Ganderbal, Jammu and Kashmir, 191131, India\\
$^{5}$Department of Physics, Gebze Technical University, 41400 Kocaeli, Turkey\\
$^{6}$Department of Physics and Astronomy, Uppsala University, 75120 Uppsala, Sweden\\
$^{7}$Department of Materials Science, School of Natural Sciences, University of Patras, GR-26504 Patras, 
Greece}

\date{\today}

\begin{abstract}

Spin gapped metals have recently emerged as promising candidates for spintronic and nanoelectronic applications, 
enabling functionalities such as sub-60~mV/dec switching, negative differential resistance, and non-local spin-valve 
effects in field-effect transistors. Realizing these functionalities, however, requires a deeper understanding of 
their magnetic behavior, which is governed by a subtle interplay between localized and itinerant magnetism. This 
interplay is particularly complex in spin gapped metallic half-Heusler compounds, whose magnetic properties remain 
largely unexplored despite previous studies of their electronic structure. In this work, we systematically investigate 
the magnetic behavior of spin gapped metallic half-Heusler compounds $XYZ$ ($X$ = Fe, Co, Ni, Rh, Ir, Pd, Pt; $Y$ = 
Ti, V, Zr, Hf, Nb, Ta; $Z$ = In, Sn, Sb), revealing clear trends. Co- and Ni-based compounds predominantly exhibit 
itinerant magnetism, whereas Ti-, V-, and Fe-based systems may host localized moments, itinerant moments, or a 
coexistence of both. To uncover the origin of magnetism, we apply the Stoner model, with the Stoner parameter $I$ 
estimated from Coulomb interaction parameters (Hubbard $U$ and Hund’s exchange $J$) computed using the constrained 
random phase approximation (cRPA). Our analysis shows that compounds not satisfying the Stoner criterion tend to remain non-magnetic.  On the contrary  compounds, which satisfy
the Stoner criterion, generally exhibit 
magnetic ordering highlighting the crucial role of electronic 
correlations and band structure effects in the emergence of magnetism. For compounds with magnetic ground states, we 
compute Heisenberg exchange parameters, estimate Curie temperatures ($T_\mathrm{C}$), and analyze spin-wave properties, 
including magnon dispersions and stiffness constants. These results provide microscopic insight into the magnetism of 
spin-gapped metallic half-Heuslers and establish a predictive framework for designing spintronic materials with tailored 
magnetic properties.

\end{abstract}

\maketitle

\section{Introduction}\label{sec1}


Heusler compounds constitute a remarkably versatile class of materials that exhibit a broad spectrum of exotic electronic and 
magnetic properties, positioning them as a central platform in the design of functional quantum materials. Originally discovered 
several decades ago, these intermetallics have recently attracted renewed interest due to their potential in spintronics, 
magnetoelectronics, and energy-related applications~\cite{Hirohata2020,Tavares2023,Chatterjee2022}. Their diverse functionalities 
arise from intricate interactions among the valence $d$-orbitals of transition metal atoms, which can be finely tuned through 
compositional and structural modifications~\cite{Graf2011,Faleev2017c,gao2019high,Ma2017,Marathe2023,Sanvito2017}. Among the 
most celebrated electronic properties of Heusler compounds is half-metallicity, in which one spin channel is metallic while 
the other is insulating, resulting in 100\% spin polarization at the Fermi level—an ideal characteristic for spintronic devices. 
A related and equally intriguing electronic structure is that of a spin gapless semiconductor, defined by a zero gap in 
one spin channel and a finite gap in the other, enabling high carrier mobility and efficient spin transport~\cite{Graf2011,Ouardi2013}. 
Certain Heusler compounds also exhibit magnetic semiconducting behavior, characterized by a semiconducting electronic structure 
combined with long-range magnetic order, which is particularly promising for spin-filtering and magneto-optical applications \cite{Spin-filter,Spin-filter2}. 
More recently, a number of Heusler compounds have been shown to host topological properties, including nontrivial surface states 
and Weyl nodes, driven by band inversions and symmetry-protected degeneracies~\cite{topo1,topo2,topo3,topo4,topo5}. 
These features position Heusler compounds at the forefront of research into quantum materials with multifunctional properties.

Among the broad family of Heusler compounds, a particularly interesting subset is formed by the 
18-valence-electron half-Heusler (or semi-Heusler) systems, which are best known for their 
semiconducting behavior and exceptional thermoelectric performance at elevated temperatures. 
These materials, including well-studied examples such as CoTiSb, FeVSb, and NiTiSn, 
have long served as model systems for exploring structure-property relationships in Heusler compounds~\cite{Ma2017,Jung2000,PIERRE1994,Tobola2000,Ouardi2012,Mokhtari2018,Emel2023,Emel2023b,Xu2013}. 
More recently, {\c{S}}a{\c{s}}{\i}o{\u{g}}lu and collaborators identified that certain 
half-Heusler compounds with one or two valence electrons more or less than these semiconductors 
are in fact non-magnetic gapped metals, characterized by an energy gap just below or above the 
Fermi level~\cite{Sasioglu2025}. Since the Fermi level intersects either the valence or conduction 
band, these gapped metals behave similarly to conventional doped semiconductors, where transport 
is dominated by holes or electrons, respectively. Remarkably, as shown in Ref.~\onlinecite{Sasioglu2025}, 
many of the studied compounds fall into a new class of materials termed \textit{spin gapped metals}. 
These spin gapped metallic Heusler compounds typically possess 16, 17, 19, or 20 valence electrons 
per formula unit and feature a spin-dependent energy gap near the Fermi level. Depending on the 
alignment of the Fermi level relative to the spin-resolved energy gaps, each spin channel can exhibit 
an intrinsic $p$- or $n$-type character, analogous to doped magnetic semiconductors \cite{Sato2010,Lei2022,Tu2014,Kroth2006}.
This intrinsic carrier-type asymmetry in spin channels makes spin gapped metals particularly 
promising for spintronic applications, as they eliminate the need for extrinsic doping and avoid 
associated issues such as disorder and phase separation. Moreover, recent theoretical proposals 
demonstrate that spin gapped metals can serve as efficient electrode materials in multifunctional 
field-effect transistors, enabling device functionalities such as sub-60 mV/dec switching, 
negative differential resistance, and non-local giant magnetoresistance \cite{multifunctional2024}. These features make spin gapped metals attractive candidates 
for future low-power, logic-in-memory, and multivalued logic devices beyond the limits of 
conventional CMOS technology.

\begin{figure*}[t]
\begin{center}
\includegraphics[scale=0.16]{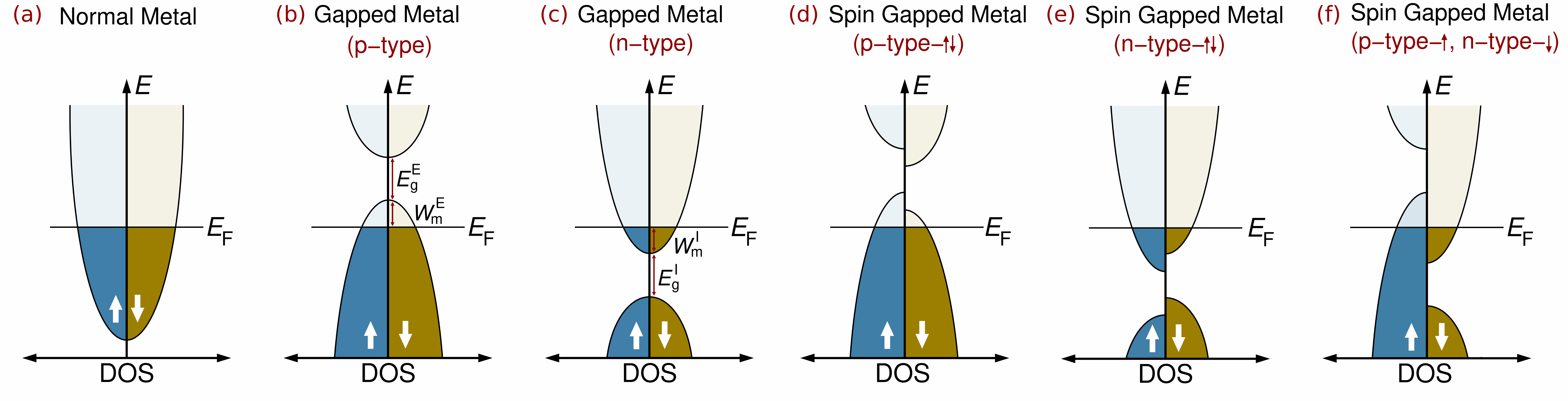}
\end{center}
\vspace*{-0.5cm} 
\caption{Schematic representation of the density of states (DOS) of a 
normal metal (a), gapped metals (b-c), and spin gapped metals (d-f). 
The arrows represent the two possible spin directions. The horizontal line 
depicts the Fermi level $E_\mathrm{F}$.} 
\label{fig1}
\end{figure*}

Despite this promising potential, the underlying magnetic behavior of spin gapped metals 
remains insufficiently understood. Although Ref.~\onlinecite{Sasioglu2025} introduced the 
concept of spin gapped metals using half-Heusler compounds as prototype materials through 
\textit{ab initio} density functional theory calculations, several aspects of their behavior 
remain unexplored. For example, some Co- and Ni-based compounds unexpectedly adopt non-magnetic 
ground states and behave as ordinary gapped metals, whereas others display magnetic 
ordering~\cite{Sasioglu2025}. These observations raise fundamental questions about the 
mechanisms that govern magnetism in these materials: Are their properties driven by 
conventional exchange interactions, or do they stem from more subtle electronic instabilities? 
Addressing these questions is essential for understanding the microscopic origin of magnetism 
in spin gapped metals and for advancing their use in spintronic applications.

In this work, we address these questions by conducting a systematic investigation of half-Heusler 
compounds incorporating a wide range of transition metals, including Ti, V, Fe, Co, Ni, Rh, Ir,
Pd, and Pt most of which were initially identified as spin gapped metals in 
Ref.~\onlinecite{Sasioglu2025}. By analyzing their magnetic moments, exchange interactions, 
and electronic structures, we uncover distinct trends in their magnetic 
behavior. Our results show that Co- and Ni-based compounds predominantly exhibit itinerant 
magnetism, whereas Ti-, V-, and Fe-based systems can display localized or itinerant moments, 
or a coexistence of both. To understand the origin of this diversity, we employ a combination 
of density functional theory (DFT), the constrained random phase approximation (cRPA), and 
the Stoner model. This integrated approach allows us to quantify electronic correlations, evaluate 
magnetic instabilities, and establish a predictive framework for magnetism in spin gapped metallic
half-Heusler compounds. Our findings not only provide microscopic insights into the mechanisms 
governing magnetism in these systems but also lay the groundwork for designing spintronic 
materials with tailored magnetic properties.

The remainder of this paper is organized as follows. In Sec.~\ref{sec2}, we provide an overview 
of the materials studied in this work, based primarily on the spin gapped metallic half-Heusler 
compounds identified in Ref.~\onlinecite{Sasioglu2025}. Section~\ref{sec3} describes the 
first-principles electronic structure methods employed, along with the theoretical models 
and approximations used to analyze magnetism and electronic correlations. In Sec.~\ref{sec4}, 
we present and discuss our results in detail. Section~\ref{sec4a} focuses on the nature of 
magnetism in each compound, identifying whether the moments are itinerant, localized, or 
coexisting. In Sec.~\ref{sec4b}, we briefly discuss the on-site effective Coulomb interaction 
parameters, which are subsequently used in Sec.~\ref{sec4c} to compute the Stoner parameter 
and evaluate the tendency of the paramagnetic state toward ferromagnetic instability. Section~\ref{sec4d} analyzes the stability of the magnetic ground state through calculated exchange constants, and 
further estimates Curie temperatures and spin-wave properties. Finally, Sec.~\ref{sec5} summarizes 
the main findings and outlines the broader implications of our work.

\section{Spin Gapped Metals}\label{sec2}

To lay the groundwork for understanding the magnetic behavior of spin gapped half-Heusler compounds, 
we begin by examining their characteristic electronic structure features. Fig.~\ref{fig1} schematically
illustrates the density of states (DOS) for normal metals, gapped metals, and spin gapped metals. In 
normal metals, the Fermi level intersects both spin channels of the band structure, and no energy gap 
exists in its vicinity. In contrast, gapped metals exhibit a gap near the Fermi level. Unlike 
semiconductors, where the Fermi level lies within the gap, in gapped metals the Fermi level either
intersects the valence band—resulting in $p$-type behavior  or the conduction band yielding $n$-type 
behavior as shown in panels (b) and (c) of Fig.~\ref{fig1}. In the $p$-type case, holes are available 
for transport, while in the $n$-type case, electrons are available, analogous to doped semiconductors. 
In spin gapped metals, the DOS becomes spin-dependent due to magnetic ordering, and the two spin 
channels exhibit distinct energy gaps. The position of these gaps relative to the Fermi level determines
the carrier type for each spin channel. Depending on this alignment, both spin channels can display 
$p$-type or $n$-type behavior [panels (d) and (e)], or a mixed case can arise [panel (f)], in which 
one spin channel is $p$-type while the other is $n$-type. To characterize these systems in more detail, 
we define four key electronic structure parameters, as depicted in Fig.~\ref{fig1}: the internal band 
gap ($E_{\mathrm{g}}^{\mathrm{I}}$), the external band gap ($E_{\mathrm{g}}^{\mathrm{E}}$), and the 
internal and external metallic bandwidths, $W_{\mathrm{m}}^{\mathrm{I}}$ and 
$W_{\mathrm{m}}^{\mathrm{E}}$, respectively. $W_{\mathrm{m}}^{\mathrm{E}}$ denotes the energy 
difference between the Fermi level and the valence band maximum for $p$-type gapped or 
spin gapped metals, while $W_{\mathrm{m}}^{\mathrm{I}}$ is the energy difference between the 
conduction band minimum and the Fermi level in $n$-type systems.

Based on the electronic classification outlined above, we now turn to the specific set 
of spin gapped half-Heusler compounds investigated in this study. In Ref.~\onlinecite{Sasioglu2025}, 
we identified spin gapped metals among half-Heusler compounds with 16, 17, and 19 valence electrons
by performing a high-throughput screening using the Open Quantum Materials Database (OQMD)~\cite{oqmd,Saal2013,Kirklin2015}. The selection criteria for candidate compounds were 
twofold. First, the formation energy $E_\text{form}$ had to be negative to ensure thermodynamic 
stability. Second, the convex hull distance $\Delta E_\text{con}$—the energy difference between 
the considered structure and the most stable phase or mixture of phases—was required to be less 
than 0.2~eV/atom, a threshold that supports the experimental feasibility of synthesizing metastable
compounds. Table~\ref{table1} summarizes all spin gapped metals studied in 
Ref.~\onlinecite{Sasioglu2025}, excluding CuVSb due to its relatively large convex hull distance.
In this work, we expand that dataset by including five additional spin gapped half-Heusler compounds:
FeZrSn, FeHfSn, NiTiIn, PdTiIn, and IrVSb. In total, we investigate twenty-four compounds, and 
their corresponding lattice constants, obtained from OQMD, are listed in Table~\ref{table1}.

\begin{table*}[t]
\caption{\label{table1}
Lattice constants ($a_0$), total number of valence electrons ($Z_T$), 
spin-gap type in each spin channel (NM stands for normal-metallic 
behavior, SC for semiconducting behavior and the arrows depict the 
spin-up and spin-down electronic band structures), sublattice and 
total magnetic moments for both ferro-/ferri-magnetic (FM) and 
antiferromagnetic (AFM) configurations, spin polarization at the 
Fermi level (see text for definition), spin-wave stiffness constant 
($D$), and calculated Curie temperatures ($T_\mathrm{C}$) for the 
studied compounds. The lattice constants $a_0$ are taken from the Open Quantum Materials Database~\cite{oqmd,Saal2013,Kirklin2015}.}
\begin{ruledtabular}
\begin{tabular}{lcrlrrrrllcl}
 &   & & & \multicolumn{3}{c}{FM (001)} & \multicolumn{2}{c}{AFM (111) } & & & \\  \cline{5-7} \cline{8-9} 
Compound & $a_0$ & Z$_T$ & Spin gap type & m$_X$ & m$_{Y}$ & m$_\text{total}$ & m$_X$ & m$_{Y}$ & SP & D & $T_\mathrm{C}$ \\
$XYZ$   & ({\AA}) & & & ($\mu_B$) & ($\mu_B$) &  ($\mu_B$) & ($\mu_B$) & ($\mu_B$)   & (\%) &  (meV\AA$^2$) & (K)  \\ \hline
FeZrSn & 6.24 & 16 & p-type-$\uparrow$/p-type-$\downarrow$ & -2.03  & 0.18  & -1.90  & -1.86  & 0.33  & 64  & 227  & 151  \\ 
FeHfSn & 6.19 & 16 & p-type-$\uparrow$/p-type-$\downarrow$ & -1.86  & 0.17  & -1.74  & -1.65  & 0.26  & 28  & 169  & 150 \\ 
FeTiSb & 5.94 & 17 & p-type-$\uparrow$/p-type-$\downarrow$ & -1.45  & 0.53  & -0.95  & -0.99  & 0.23  & 68  & 706  & 317 \\ 
FeZrSb & 6.15 & 17 & p-type-$\uparrow$/SC-$\downarrow$     & -1.34  & 0.34  & -1.00  & -1.08  & 0.16  & 100 & 738  & 274 \\ 
FeHfSb & 6.11 & 17 & p-type-$\uparrow$/SC-$\downarrow$     & -1.26  & 0.27  & -1.00  & -0.93  & 0.12  & 100 & 857  & 276 \\ 
FeVSn  & 5.87 & 17 & NM-$\uparrow$/p-type-$\downarrow$     & -1.85  & 1.03  & -0.88  & -1.40  & 0.87  & 36  & 383  & 532 \\ 
FeNbSn & 6.00 & 17 & p-type-$\uparrow$/SC-$\downarrow$     & -1.38  & 0.40  & -1.00  & -1.05  & 0.23  & 100 & 734  & 266 \\ 
FeTaSn & 5.99 & 17 & p-type-$\uparrow$/SC-$\downarrow$     & -1.29  & 0.32  & -1.00  & -0.88  & 0.18  & 100 & 810  & 282 \\ 
CoTiSn & 5.93 & 17 & p-type-$\uparrow$/p-type-$\downarrow$ & -0.42  & -0.45 & -0.94  &  0.00  & 0.00  & 74  & 669  & 56  \\ 
CoZrSn & 6.15 & 17 & p-type-$\uparrow$/p-type-$\downarrow$ & -0.67  & -0.18 & -0.95  & -0.27  & 0.01  & 81  & 936  & 126 \\ 
CoHfSn & 6.11 & 17 & p-type-$\uparrow$/p-type-$\downarrow$ & -0.55  & -0.15 & -0.79  & 0.00   & 0.00  & 66  & 897  & 108 \\ 
RhTiSn & 6.17 & 17 & p-type-$\uparrow$/p-type-$\downarrow$ & -0.07  & -0.73 & -0.87  & 0.00   & 0.00  & 70  & 226  & 131 \\ 
IrTiSn & 6.20 & 17 & p-type-$\uparrow$/p-type-$\downarrow$ & -0.08  & -0.61 & -0.76  & 0.00   & 0.00  & 33  & 165  & 95  \\ 
NiTiIn & 5.99 & 17 & p-type-$\uparrow$/p-type-$\downarrow$ & -0.04  & -0.81 & -0.97  & 0.00   & 0.00  & 97  & 533  & 141 \\ 
NiZrIn & 6.22 & 17 & p-type-$\uparrow$/p-type-$\downarrow$ & -0.11  & -0.31 & -0.55  & 0.00   & 0.00  & 53  & 177  & 23  \\ 
PdTiIn & 6.23 & 17 & p-type-$\uparrow$/SC-$\downarrow$     & -0.03  & -0.87 & -0.99  & -0.01  & -0.27 & 99  & 713  & 157 \\ 
PtTiIn & 6.24 & 17 & p-type-$\uparrow$/SC-$\downarrow$     & -0.04  & -0.84 & -1.00  & 0.01   & -0.19 & 100 & 769  & 189 \\ 
CoVSb  & 5.81 & 19 & n-type-$\uparrow$/SC-$\downarrow$     & -0.33  &  1.41 &  1.00  & 0.11   & 1.31  & 100 & 590  & 419 \\ 
RhVSb  & 6.07 & 19 & n-type-$\uparrow$/SC-$\downarrow$     & -0.21  &  1.33 &  1.00  & -0.08  & 1.64  & 100 & 484  & 228 \\ 
IrVSb  & 6.07 & 19 & n-type-$\uparrow$/SC-$\downarrow$     & -0.21  &  1.26 &  1.00  & -0.08  & 1.30  & 100 & 498  & 283 \\ 
PdTiSb & 6.24 & 19 & n-type-$\uparrow$/n-type-$\downarrow$ & -0.04  &  0.98 &  0.89  & -0.02  & 0.83  & 87  & 383  & 194 \\ 
PtTiSb & 6.26 & 19 & n-type-$\uparrow$/n-type-$\downarrow$ & -0.05  &  1.05 &  0.99  & -0.03  & 0.76  & 93  & 755  & 318 \\ 
NiVSn  & 5.87 & 19 & n-type-$\uparrow$/SC-$\downarrow$     & -0.03  &  1.15 &  1.00  & -0.03  & 1.77  & 100 & 250  & 109 \\ 
NiVSb  & 5.88 & 20 & n-type-$\uparrow$/SC-$\downarrow$     & 0.03   &  2.12 &  2.00  & 0.01   & 2.29  & 100 & 561  & 684 \\ 
\end{tabular}
\end{ruledtabular}
\end{table*}

Beyond structural stability, previous studies have also revealed systematic trends in the electronic 
properties of these compounds, which provide important context for the magnetic behavior analyzed in 
this work. In Ref.~\onlinecite{Sasioglu2025}, a detailed analysis of the spin-resolved electronic
band structures was provided for each compound. For reasons of completeness   we have included in Table \ref{table1} the 
spin-gap type of the band structure for both spin channels 
for the compounds under study. A clear trend was identified: compounds with fewer 
than 18 valence electrons generally exhibit $p$-type spin gapped behavior, whereas those with more 
than 18 valence electrons tend to be $n$-type spin gapped metals. The nature of spin polarization at 
the Fermi level further distinguishes the electronic character. The spin polarization (SP) is defined 
as the difference between the spin-up and spin-down DOS at the Fermi level divided by the total DOS. 
If the SP is 100\%, one spin channel (typically spin-down) exhibits a semiconducting character. When 
SP is less than 100\%, both spin channels contribute to conduction and display either $p$- or $n$-type
metallic behavior. Table~\ref{table1} lists also the SP values for all compounds considered in this study. 
Most compounds with 16 or 17 valence electrons exhibit $p$-type spin gapped character in both spin 
channels. In contrast, among the 19- or 20-valence-electron systems, only PdTiSb and PtTiSb are 
$n$-type in both spin channels; the remaining compounds show $n$-type behavior in one spin direction, 
while the other exhibits semiconducting characteristics.

\begin{figure*}
\begin{center}
\includegraphics[scale=0.11]{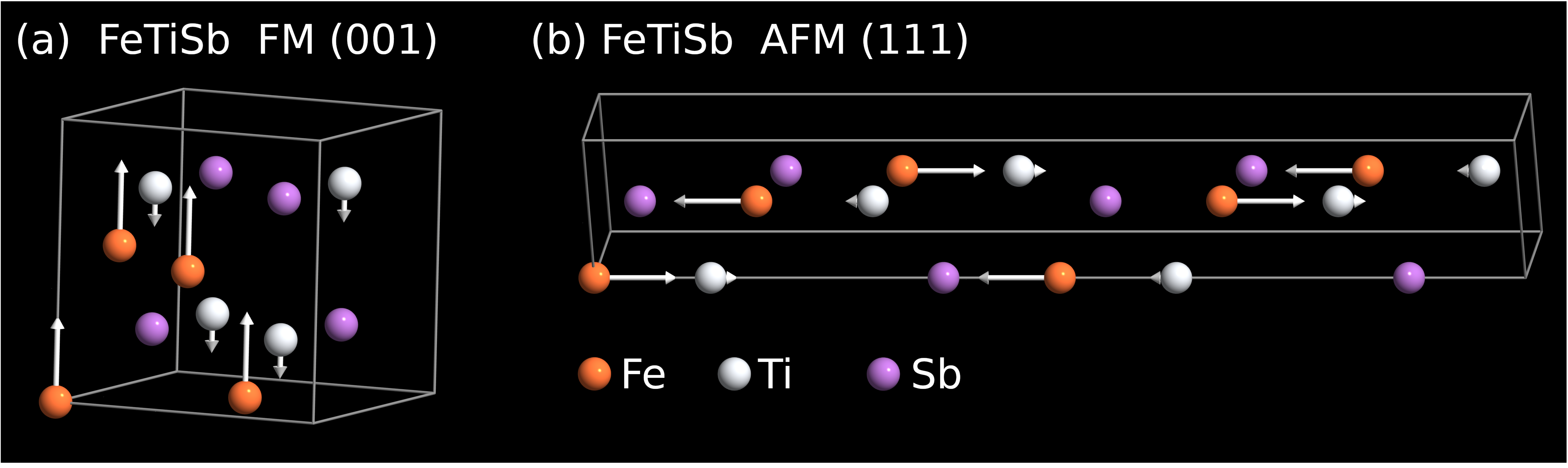}
\end{center}
\vspace*{-0.3cm} 
\caption{Schematic representation of the unit cells used for ferro-/ferri-magnetic (FM) and antiferromagnetic 
(AFM) calculations, illustrated using the FeTiSb compound as an example. Panel (a) shows the conventional 
unit cell employed for FM calculations, where atomic spin magnetic moments (indicated by arrows) are aligned 
parallel to the [001] crystallographic direction. Panel (b) depicts the doubled unit cell constructed along 
the [111] direction for AFM calculations, in which the magnetic moments are oriented antiparallel along [111].
The length of each arrow is proportional to the magnitude of the calculated atomic spin magnetic moment.}
\label{fig2}
\end{figure*}

To complement the electronic structure perspective, it is also important to consider the 
crystallographic framework of the spin gapped metals studied in this work, as crystal symmetry 
plays a crucial role in governing magnetic interactions and possible magnetic ordering. All 
bulk half-Heusler compounds with the general formula $XYZ$ crystallize in the cubic $C1_\mathrm{b}$
structure, which belongs to the $F\overline{4}3m$ space group. A schematic representation of this 
lattice is shown in Fig.~\ref{fig2}(a) for the representative case of FeTiSb. The large cubic 
unit cell shown contains four primitive unit cells and can be viewed as a face-centered cubic 
(fcc) lattice with a four-site basis. In Wyckoff coordinates, the A site at $(0\,0\,0)$ is occupied 
by Fe atoms, the B site at $(\frac{1}{4}\,\frac{1}{4}\,\frac{1}{4})$ by Ti atoms, and the D site 
at $(\frac{3}{4}\,\frac{3}{4}\,\frac{3}{4})$ by Sb atoms. The C site at $(\frac{1}{2}\,\frac{1}{2}\,\frac{1}
{2})$ remains unoccupied—a characteristic feature of the $C1_\mathrm{b}$ structure. 
From a symmetry standpoint, both the Fe and the vacant C sites are located at the centers of smaller 
cubes, each surrounded by four Ti and four Sb atoms as nearest neighbors at the cube corners. 
Conversely, each Ti or Sb atom is centered within a cube defined by four Fe atoms and four vacant 
C sites. Fig.~\ref{fig2}(b) shows the doubled unit cell constructed along the [111] direction, which 
we employ for our antiferromagnetic (AFM) calculations. The specific AFM configuration and its 
implications for the magnetic behavior will be discussed in Sec.~\ref{sec4}.

\section{Computational Details} \label{sec3}

Having established the electronic and structural characteristics of spin gapped metsllic half-Heusler
compounds, we now turn to the computational framework used to investigate their magnetic behavior. In 
this study, we employ three different \textit{ab initio} electronic structure methods to investigate 
the ground-state properties of spin gapped metals. Our extensive cross-validation tests (not shown) 
confirm that all three methods yield nearly identical results for the systems under study when the 
same exchange-correlation functional is used. This is consistent with the findings of 
Ref.~\onlinecite{Lejaeghere2016}, which provides a comprehensive comparison of the accuracy of a wide 
range of first-principles electronic structure approaches. Throughout 
this work, we use the Perdew–Burke–Ernzerhof (PBE) parameterization 
of the generalized gradient approximation (GGA) for the 
exchange-correlation functional~\cite{Perdew1996}, consistent with our earlier study on spin gapped
metals~\cite{Sasioglu2025}. The PBE functional is known to deliver reliable results for Heusler compounds,
even in comparison to more sophisticated functionals~\cite{Meinert2013}. The rationale for using three
different methods lies in the flexibility they offer for post-processing and analyzing different magnetic 
properties. Each method provides complementary capabilities, allowing us to perform an in-depth 
investigation of the electronic and magnetic behavior of spin gapped half-Heusler compounds.

The first \textit{ab initio} electronic structure method we employ is the \textsc{QuantumATK} software 
package~\cite{QuantumATK,QuantumATKb}. This approach uses linear combinations of atomic orbitals (LCAO) as a 
basis set, combined with norm-conserving PseudoDojo pseudopotentials~\cite{VanSetten2018}. For ground-state 
calculations of the bulk compounds, we use a $16 \times 16 \times 16$ Monkhorst–Pack $\mathbf{k}$-point 
grid~\cite{Monkhorst1976}.  In addition to ground-state calculations, the \textsc{QuantumATK} framework 
is used to compute magnetic interactions and excitations, making it central to our analysis of spin dynamics.
Specifically, we calculate the Heisenberg exchange parameters using the Liechtenstein-Katsnelson-Antropov-Gubanov (LKAG) 
formalism~\cite{liechtenstein1987local}, following the procedure outlined 
in Ref.~\onlinecite{Emel2023}. These exchange constants serve as the foundation for two key analyses: estimation 
of Curie temperatures and evaluation of spin-wave properties. To estimate the Curie temperatures ($T_\mathrm{C}$), 
the computed exchange parameters are used within the mean-field approximation. For the spin-wave analysis, we 
calculate the magnon dispersions and extract spin-wave stiffness constants based on the formalism developed 
in Ref.~\onlinecite{He2021}. This method, tailored for multi-sublattice magnetic systems, generalizes earlier 
work by Pajda \textit{et al.}~\cite{Pajda2001}, originally formulated for single-sublattice magnets. Together, 
these calculations provide a comprehensive picture of the magnetic exchange interactions and excitation spectra, 
offering valuable insights into the potential of spin gapped metals for high-temperature spintronic applications.

The second \textit{ab initio} method employed in this work is the full-potential linearized augmented 
plane wave (FLAPW) approach, as implemented in the \textsc{FLEUR} code~\cite{FLEUR,FLEUR2}. We first 
perform non-magnetic ground-state calculations using \textsc{FLEUR} and then use the \textsc{SPEX} 
code~\cite{SPEX,SPEX2} to compute the effective Coulomb interaction parameters within the framework 
of the constrained random phase approximation (cRPA). The cRPA method is a state-of-the-art approach 
for determining material-specific interaction parameters such as the Hubbard $U$ and Hund's exchange $J$. 
It has been successfully applied to a variety of Heusler systems, including half-metallic~\cite{Sasioglu2013} 
and Mn-based full Heusler compounds~\cite{Tas2022}. For further details on the cRPA methodology in Heusler 
compounds, we refer the reader to Refs.~\onlinecite{Sasioglu2013,Tas2022}. The obtained $U$ and $J$ 
values are then used to compute the Stoner parameter, defined as $I = (U + 6J)/5$~\cite{Stollhoff1990}.
Within the Stoner model, this parameter plays a central role in predicting ferromagnetic instability through 
the criterion $I \cdot N(E_F) > 1$, where $N(E_F)$ denotes the density of states at the Fermi level.

Finally, we employ the all-electron \textit{ab initio} full-potential nonorthogonal local-orbital minimum-basis 
band structure method (FPLO)~\cite{FPLO,FPLO2}. This approach allows for an accurate and efficient treatment of 
the electronic structure, particularly near the atomic cores. In this work, we use the FPLO results primarily to 
compute and visualize the charge density isosurfaces, which will be discussed in detail in the following sections.
Together, these three complementary methods enable a detailed and cross-validated analysis of the electronic, 
magnetic, and correlation effects in spin gapped metallic half-Heusler compounds.

\section{Results and Discussion}\label{sec4}

\subsection{Itinerant versus localized magnetism} \label{sec4a}

To gain insight into the nature of magnetism in 16-, 17-, 19-, and 20-valence-electron half-Heusler 
compounds, we computed the magnetic moments for both ferromagnetic/ferrimagnetic (FM) and antiferromagnetic 
(AFM) configurations (Table~\ref{table1}). For the AFM configuration, we considered a layered [111]-oriented 
ordering, where spin magnetic moments alternate in direction between neighboring (111) planes. The unit 
cells used for FM and AFM calculations are illustrated in Fig.~\ref{fig2}, with FeTiSb shown as a representative 
example. In these schematics, the arrow directions indicate spin magnetic moments orientation, while their lengths represent 
the magnitude of the magnetic moments. This computational setup provides a straightforward yet effective means
to distinguish between localized and itinerant magnetism: in systems with localized moments, magnetic moments 
tends to persist even in the AFM configuration, whereas itinerant moments are typically suppressed or vanish 
upon switching from FM to AFM alignment. The results summarized in Table~\ref{table1} reveal distinct trends 
across the different valence electron counts, highlighting a strong correlation between electronic structure 
and moment stability. Specifically, Fe-based 16- and 17-electron compounds exhibit relatively stable local 
moments in both FM and AFM states. In contrast, several 17-electron systems—particularly those based on Co, 
Rh, and Ni—display significant suppression or even complete collapse of the magnetic moment in the AFM 
configuration, consistent with itinerant behavior. Meanwhile, 19- and 20-electron compounds maintain robust 
magnetism with minimal variation between FM and AFM states, underscoring the role of exchange interactions 
in stabilizing local moments in these systems.

To understand the origin of these trends, we now examine the underlying electronic mechanisms that govern 
moment stability across different compound families. The collapse of magnetic moments in certain 
17-valence-electron compounds can be attributed to the interplay between exchange splitting and electronic 
hybridization. In Co- and Rh-based systems, the relatively large $d$-bandwidth leads to stronger hybridization, 
promoting more itinerant magnetic behavior. As a consequence, AFM ordering significantly modifies the DOS 
at the Fermi level, suppresses exchange splitting, and destabilizes local moments. This trend is evident in 
compounds such as CoTiSn and CoHfSn, where the magnetic moment completely vanishes in the AFM 
configuration. In contrast, Fe-based 16- and 17-electron compounds retain sizable moments even in 
the AFM state, indicating a more localized character of magnetism on Fe atoms, which is less 
sensitive to spin-ordering changes. For 19- and 20-electron compounds, such as CoVSb, RhVSb, and NiVSb, 
the magnetic moments in FM and AFM configurations remain largely unchanged, suggesting a robust exchange 
interaction that stabilizes magnetism regardless of magnetic ordering. This is particularly evident in 
NiVSb, where a sizable moment of approximately $2.12 \, \mu_B$ persists in the AFM state.

\begin{figure}
\begin{center}
\includegraphics[width=\columnwidth]{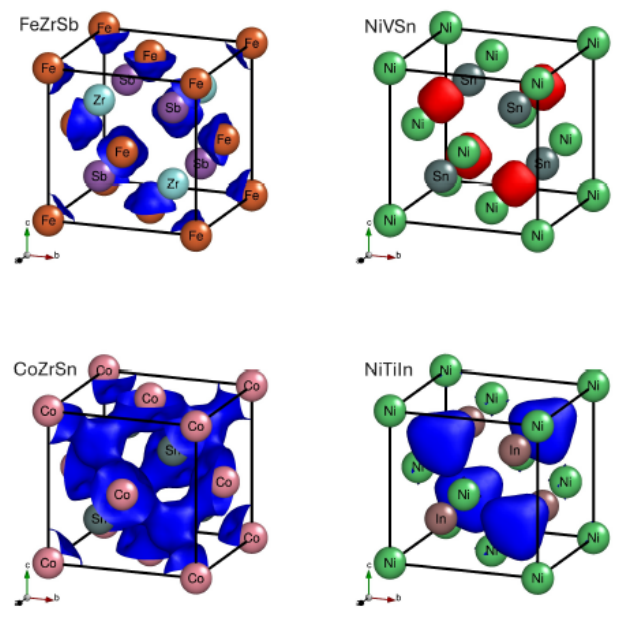}
\end{center}
\vspace*{-0.7cm} 
\caption{Spin density isosurfaces (defined as the difference between spin-up and spin-down charge densities)
for four representative compounds. FeZrSb and NiVSn exhibit localized magnetic moments centered on the Fe and V 
atoms, respectively. In contrast, CoZrSn and NiTiIn display itinerant magnetic behavior, with spin polarization 
primarily associated with the Co and Ni atoms. Positive and negative values of the spin density are represented 
by red and blue isosurfaces, respectively.}
\label{fig3}
\end{figure}

To further support these interpretations, we analyze the real-space spin density distributions, which offer 
direct insight into the degree of magnetic moment localization. Fig.~\ref{fig3} presents the spin density 
— defined as the difference between spin-up and spin-down charge densities — for four representative 
compounds: FeZrSb, CoZrSn, NiVSn, and NiTiSn. In FeZrSb, which exhibits localized magnetism, the spin density 
is primarily concentrated around the Fe atoms. In contrast, CoZrSn displays itinerant magnetic behavior, with 
the spin density distributed more broadly around both Co and Zr sites. NiVSn and NiTiSn illustrate contrasting 
cases: while NiVSn shows localized moments centered on the V atoms, NiTiSn exhibits more delocalized spin density 
around Ti, consistent with itinerant magnetism. The spin-density isosurfaces are shown in blue and red, 
representing negative and positive values of the same magnitude, respectively. In FeZrSb, the isosurface forms 
a compact, sphere-like shape tightly surrounding the Fe atoms, indicative of localized magnetism. For CoZrSn, 
the isosurface forms an extended, hollow network encompassing both Co and Zr atoms—a hallmark of itinerant behavior. 
In NiVSn and NiTiSn, the isosurfaces around the V and Ti atoms are also roughly spherical, but their radii differ 
significantly: the larger isosurface radius in NiTiSn reflects its itinerant character, whereas the smaller, more 
localized red spheres in NiVSn resemble those of FeZrSb, indicating localized magnetism.

Overall, the observed trends in magnetic moment behavior reflect a delicate interplay between exchange splitting, 
electronic hybridization, and spin ordering in half-Heusler compounds. The pronounced collapse of magnetic moments 
in certain Co- and Rh-based systems underscores the role of itinerant magnetism, whereas the robustness of magnetism 
in Fe- and V-based compounds points to a more localized character. These insights deepen our understanding of 
magnetism in spin gapped metallic half-Heuslers and provide a microscopic foundation for tailoring their electronic
and magnetic properties in future spintronic applications.

\begin{table*}[t]
\caption{\label{table2}
Effective Coulomb interaction parameters (Hubbard $U$ and Hund exchange $J$) for the valence $d$-orbitals of transition metal 
atoms in the studied half-Heusler compounds, along with the estimated Stoner parameter $I$ and the total density of states at the 
Fermi level $N(E_F)$ obtained from non-spin-polarized calculations. The product $I\cdot N(E_F)$ is presented as a direct measure of 
the Stoner instability, with an additional renormalized value $\alpha \cdot I \cdot N(E_F)$, where $\alpha = 0.6$ accounts for many-body 
effects (see text for details). The slash separates the values for the two transition metal atoms $X$ and $Y$ in the chemical formula 
of the compound. This dataset allows for a sublattice-resolved application of the Stoner criterion, distinguishing cases where both, 
one, or neither sublattice satisfies the condition for spontaneous magnetization.}
\begin{ruledtabular}
\begin{tabular}{lccccccc}
$XYZ$ & Orbitals & $U$(eV) & $J$(eV) & $I$(eV) & N(E$_F$)(eV$^{-1}$) & $I\cdot N(E_F)$ & $\alpha\cdot I \cdot N(E_F)$  \\
\hline
\multicolumn{8}{c}{spin gapped metals} \\
FeZrSn &  3$d$/4$d$ & 1.93/1.50 & 0.88/0.39 & 1.44/0.76  & 1.50/1.21   &  2.16/0.92 & 1.30/0.55    \\
FeHfSn &  3$d$/5$d$ & 2.08/1.40  & 0.89/0.36 & 1.48/0.71  & 1.28/0.76 &  1.84/0.54  & 1.14/0.32   \\
FeTiSb &  3$d$/3$d$ & 2.42/2.65 & 0.85/0.61  & 1.50/1.27  & 1.72/1.42  &  2.58/1.80 & 1.55/1.08  \\     
FeZrSb &  3$d$/4$d$ & 2.73/1.75 & 0.90/0.37 & 1.62/0.79  & 1.75/0.62  &  2.84/0.49 & 1.70/0.29  \\      
FeHfSb &  3$d$/5$d$ & 2.91/1.52 & 0.91/0.33 & 1.67/0.71  & 1.51/0.89   &  2.52/0.63 & 1.51/0.38  \\    
FeVSn  &  3$d$/3$d$ &  2.13/2.67 & 0.85/0.74 & 1.45/1.42  & 1.52/0.94  &  2.20/1.33 & 1.32/0.80  \\    
FeNbSn &  3$d$/4$d$ &  2.20/1.96 & 0.89/0.47  & 1.51/0.96 & 1.50/0.53  &  2.27/0.51 & 1.36/0.31  \\      
FeTaSn &  3$d$/5$d$ &  2.28/1.76 & 0.90/0.42  & 1.54/0.86 & 1.46/0.54  &  2.25/0.46 & 1.35/0.28  \\      
CoTiSn &  3$d$/3$d$ &  2.33/2.11 & 0.94/0.63 & 1.59/1.17  & 1.07/1.45  &  1.70/1.70 & 1.02/1.02   \\       
CoZrSn &  3$d$/4$d$ &  2.12/1.45 & 0.96/0.38 & 1.57/0.75 &  1.48/1.25  &  2.32/0.94 & 1.39/0.56  \\            
CoHfSn &  3$d$/5$d$ &  2.58/1.39 & 0.98/0.36  & 1.70/0.70 & 1.15/0.92  &  1.96/0.64 & 1.17/0.39  \\     
RhTiSn &  4$d$/3$d$ &  2.17/1.78 & 0.66/0.65  & 1.22/1.13 &  0.42/1.56 &  0.51/1.76 & 0.31/1.06  \\      
IrTiSn &  5$d$/3$d$ &  1.89/1.86 & 0.58/0.65 & 1.08/1.15 &  0.39/1.49  &  0.42/1.71 & 0.25/1.03  \\   
NiTiIn &  3$d$/3$d$ &  3.09/1.47 & 1.05/0.61 & 1.88/1.03 &  0.62/2.52  &  1.17/2.60 & 0.70/1.56  \\ 
NiZrIn &  3$d$/4$d$ &  3.00/1.23 & 1.07/0.39 & 1.89/0.72  &  0.60/1.56  &  1.13/1.12 & 0.68/0.67  \\ 
PdTiIn &  4$d$/3$d$ &  2.63/1.29 & 0.72/0.62  & 1.39/1.00 &  0.20/2.12  &  0.28/2.12 & 0.17/1.27  \\  
PtTiIn &  5$d$/3$d$ &  2.28/1.41  & 0.65/0.52 & 1.23/0.90  &  0.30/2.50 &  0.37/2.25 & 0.22/1.35  \\
CoVSb  &  3$d$/3$d$ &  3.65/3.14 & 0.94/0.71 & 1.85/1.48  &  0.98/4.36 &  1.81/6.45 & 1.09/3.87  \\   
RhVSb  &  4$d$/3$d$  &  2.90/3.00 & 0.65/0.73 & 1.36/1.48 &  0.37/3.68 &  0.50/5.45 & 0.30/3.27  \\
IrVSb  &  5$d$/3$d$ &  2.62/3.12 & 0.58/0.73 & 1.22/1.50  &  0.17/2.13 &  0.21/3.20 & 0.12/1.92  \\
PdTiSb &  4$d$/3$d$ &  3.31/2.58 & 0.74/0.62 & 1.54/1.26 & 0.29/1.84   &  0.45/2.32 & 0.27/1.39 \\      
PtTiSb &  5$d$/3$d$  &  2.85/2.48 & 0.64/0.58 & 1.33/1.19 & 0.28/3.29 &  0.37/3.90 & 0.22/2.35  \\      
NiVSn  &  3$d$/3$d$ &  4.15/2.71 & 1.04/0.72 & 2.08/1.40 & 0.69/2.74 & 1.44/3.84 & 0.86/2.30  \\    
NiVSb  &  3$d$/3$d$ &  3.95/2.82 & 1.02/0.69 & 2.02/1.39 & 1.52/7.51 & 3.07/10.4 & 1.84/6.26  \\

\multicolumn{8}{c}{Gapped metals} \\
NiHfIn &  3$d$/5$d$ &  3.37/1.20 & 1.09/0.37 & 1.98/0.68  &  0.50/1.30  &  0.99/0.88 & 0.59/0.53 \\
CoNbSb &  3$d$/4$d$  &  4.05/2.20 & 0.98/0.45  & 1.98/0.98 & 0.70/1.18 & 1.39/1.16 & 0.83/0.69   \\  
CoTaSb &  3$d$/5$d$ &  4.08/1.93 & 0.98/0.39 & 1.99/0.86 &  0.56/0.76  &  1.11/0.65 & 0.67/0.39 \\    
NiTiSb &  3$d$/3$d$ &  4.26/2.95 & 1.04/0.58 & 2.10/1.29 &  0.41/1.16  &  0.86/1.50 & 0.52/0.90 \\     
NiZrSb &  3$d$/4$d$ &  4.82/1.96 & 1.08/0.36  & 2.27/0.82 & 0.27/0.47  &  0.61/0.39 & 0.37/0.23   \\     
NiHfSb &  3$d$/5$d$ &  4.91/1.69  & 1.09/0.32 & 2.30/0.73 & 0.23/0.39  &  0.53/0.29 &0.32/0.17 \\      
NiNbSn &  3$d$/4$d$ &  4.72/2.19 & 1.09/0.46  & 2.25/0.99 & 0.62/1.30 & 1.40/1.29 & 0.84/0.77 \\    
NiTaSn &  3$d$/5$d$ &  4.85/1.99 & 1.10/0.41 & 2.29/0.90 & 0.40/0.75 & 0.92/0.68 &0.55/0.41 
\end{tabular}
\end{ruledtabular}
\end{table*}

\subsection{Coulomb interaction parameters: Hubbard $U$ and Hund exchange $J$}
\label{sec4b}

To complement the analysis of magnetic moment behavior, we next examine the role of electronic correlations 
in shaping the magnetic properties of spin gapped half-Heusler compounds. In particular, we focus on the 
effective on-site Coulomb interaction parameters — Hubbard $U$ and Hund’s exchange $J$ — which serve as critical
inputs for estimating the Stoner parameter $I$ and evaluating ferromagnetic instability, as discussed in 
the next subsection. Since direct experimental determination of $U$ and $J$ is notoriously difficult and 
reliable data is limited, we computed these parameters from first principles using the constrained random 
phase approximation (cRPA) method implemented in the \textsc{SPEX} code, following the methodology outlined
in Ref.~\onlinecite{Tas2022} (see also Sec.~\ref{sec3}).  The resulting $U$ and $J$ values for all compounds 
are summarized in Table~\ref{table2}, where the slash separates the contributions associated with the $d$ 
orbitals of the two transition metal elements, $X$ and $Y$. These parameters are especially important in systems 
containing 3$d$, 4$d$, and 5$d$ transition metals, which often exhibit strong electronic correlations. In such 
cases, beyond-DFT methods like DFT+$U$ and DFT+DMFT have been shown to play an essential role in accurately 
describing the magnetic and electronic properties~\cite{Minar2011,Solovyev2008,Karlsson2010,Lechermann2006}. 
Heusler compounds, in particular, have been widely studied within this framework due to their rich 
correlation-driven physics~\cite{Shourov2021,Fischer2020}.

With this approach, we obtain a consistent and physically meaningful set of $U$ and $J$ parameters for all 
investigated compounds. The computed Hubbard $U$ values, which quantify the on-site Coulomb repulsion among 
$d$ electrons, are broadly comparable to those of the corresponding elemental transition metals, typically 
ranging between 1.5 and 5.7~eV~\cite{Sasioglu2011}. However, identifying systematic trends in the $U$ values 
listed in Table~\ref{table2} remains challenging, as $U$ is known to depend sensitively on factors such as 
crystal symmetry, $d$-orbital filling, and hybridization effects. In the spin gapped metallic half-Heuslers 
studied here, this complexity is amplified by the ternary nature of the unit cell, where hybridization between 
the $d$ orbitals of the $X$ and $Y$ atoms, as well as the influence of the $Z$ element, significantly affects
the resulting Coulomb interactions — consistent with observations in Mn-based full-Heusler systems~\cite{Tas2022}.
The calculated Hund’s exchange $J$ parameters are consistently smaller than their $U$ counterparts, typically 
remaining below 1~eV. These values are in line with those reported for both elemental transition metals~\cite{Sasioglu2011}
and other Heusler compounds~\cite{Sasioglu2013,Tas2022}, further supporting the reliability of our cRPA-based 
results. In the following subsection, we use these parameters to estimate the Stoner criterion and assess the
conditions under which ferromagnetic ordering becomes energetically favorable in spin gapped half-Heusler compounds.

\subsection{Stoner criterion and magnetic instabilities}\label{sec4c}

The Stoner model provides a fundamental framework for understanding the onset of ferromagnetism in 
transition-metal-based compounds by evaluating the instability of the non-magnetic (NM) state toward 
spontaneous spin polarization. Its predictive power is particularly relevant for systems in which 
magnetism arises from an itinerant-electron mechanism, driven by electronic structure rather than 
localized moments. As seen in our analysis of FM and AFM magnetic moments (Sec.~\ref{sec4a}) and 
electronic correlations (Sec.~\ref{sec4b}), a clear distinction emerges between compounds exhibiting 
localized magnetism—such as those based on Fe and V—and those where magnetic moments are more itinerant, 
notably in Co-, Rh-, and Ni-based systems. For the latter group, the Stoner criterion serves as an 
effective tool to assess the tendency of the NM state to develop finite spin polarization. In contrast, 
for compounds with robust local moments stabilized by strong exchange interactions and Hund’s coupling, 
the Stoner model alone is insufficient. In these cases, complementary local-moment-based approaches, 
such as Heisenberg exchange models, are required to capture the underlying magnetic behavior.

To quantitatively assess the applicability of the Stoner model to the half-Heusler compounds studied 
here, we use the effective Coulomb interaction parameters (Hubbard $U$ and Hund’s exchange $J$) obtained 
in the previous subsection to estimate the Stoner parameter $I$ (see Table~\ref{table2}). The Stoner 
parameter is computed using the mean-field expression proposed by Stollhoff \textit{et al.}, given by 
$I = (U + 6J)/5$~\cite{Stollhoff1990}. This relation, derived from the Hartree–Fock solution of the 
multi-orbital Hubbard model, has been shown to yield reliable results for transition-metal systems and 
reproduces the Stoner parameters of elemental 3$d$ materials with good accuracy. To evaluate magnetic 
stability, we also calculated the total density of states at the Fermi level, $N(E_F)$, in the non-magnetic 
(NM) state and determined the Stoner product $I \cdot N(E_F)$. According to the conventional criterion, 
a compound is predicted to be ferromagnetic when $I \cdot N(E_F) > 1$. For systems with two different 
transition-metal atoms, such as FeVSn or CoVSb, this condition must be evaluated separately for each 
sublattice, since the original Stoner model was formulated for a single magnetic species. These materials 
can be treated as two-sublattice magnets, where each sublattice corresponds to one of the transition-metal
atoms. Accordingly, the instability conditions become:
\[
I_X \cdot N_X(E_F) > 1, \quad I_Y \cdot N_Y(E_F) > 1,
\]
where $I_X$ and $I_Y$ are the Stoner parameters for the two sublattices, and $N_X(E_F)$ and $N_Y(E_F)$ 
are the corresponding projected DOS at the Fermi level. If both conditions are satisfied, magnetic moments 
are expected on both sublattices. If only one is satisfied, magnetism is likely to be localized on that 
sublattice, while the other remains non-magnetic or possesses
a small induced spin magnetic moment due to 
hybridization effects between orbitals sitting at 
neigboring atoms and transforming with the same symmetry 
(\textit{e.g.} Zr in FeZrSb).
If neither inequality holds, the system is expected 
to remain non-magnetic, in agreement with our DFT results. It is worth noting that this analysis neglects 
many-body correlation effects, which are known to reduce the effective Stoner parameter. According to 
Ref.~\onlinecite{Stollhoff1990}, such correlations can suppress $I$ by approximately 40\%. To account 
for this, we also consider a renormalized version of the criterion, $\alpha \cdot I \cdot N(E_F)$, with 
$\alpha = 0.6$, to provide a more realistic estimate of magnetic instability in correlated systems.

Our calculations reveal three distinct types of magnetic behavior among the half-Heusler compounds, 
classified according to whether the Stoner criterion is satisfied on one, both, or neither of the
transition-metal sublattices. In compounds such as FeTiSb, FeVSn, CoTiSn, and CoVSb, both transition-metal 
atoms satisfy the Stoner condition, resulting in finite magnetic moments on both sublattices. In a 
second group of systems, only one sublattice meets the criterion, leading to magnetism localized on 
a single atomic site. For example, in RhTiSn and PdTiIn, the Ti atom develops a sizable magnetic moment, 
while the Rh and Pd atoms remain non-magnetic due to their low projected DOS at the Fermi level, which 
fails to induce spin instability. For FeZrSb, only Fe atoms satisfy
the Stoner criterion presenting a large value of spin magnetic moment while Zr atoms, which do not
satisfy the Stoner criterion, present a much smaller spin magnetic
moment induced by the Fe atomic spin magnetic moments. The third group comprises materials in which neither sublattice satisfies 
the Stoner criterion, leading to a non-magnetic ground state. These compounds, which we classify as 
\textit{gapped metals}, include NiHfIn, CoNbSb, CoTaSb, NiTiSb, NiZrSb, NiHfSb, NiNbSn, and NiTaSn. 
Despite their metallic character, these systems exhibit a strongly reduced total DOS at the Fermi level, 
which suppresses magnetic instability. In all such cases, both the unrenormalized and renormalized 
Stoner products, $I \cdot N(E_F)$ and $\alpha \cdot I \cdot N(E_F)$, remain below the critical threshold
of unity. The absence of magnetism in these materials is primarily attributed to strong hybridization 
effects and electronic band structure features that lower the DOS at $E_F$. This sets them apart from 
nearly ferromagnetic metals such as Pd, where a high DOS near the Fermi level leads to strong spin 
fluctuations and exchange-enhanced paramagnetism.

Despite the widespread use of the Stoner model for predicting magnetism in transition-metal-based systems,
a fully developed multi-sublattice extension remains lacking. Lichtenstein \textit{et al.}~\cite{lichtenstein2013magnetism} proposed an advanced theoretical framework that combines Stoner, Heisenberg, and Hubbard concepts through 
a site- and orbital-resolved spin-fluctuation theory within the LDA+DMFT formalism. While this approach does 
not explicitly define a sublattice-resolved Stoner criterion in the conventional sense, it provides strong 
conceptual justification for evaluating magnetic instabilities on inequivalent atomic sites individually.
Our results demonstrate that such a sublattice-based application of the Stoner criterion is highly effective 
in capturing the diverse magnetic behavior observed in half-Heusler compounds, particularly in systems exhibiting
a coexistence of localized and itinerant magnetism. Nevertheless, for materials with strong local moments or 
complex magnetic interactions, additional theoretical treatments—such as Heisenberg exchange modeling or beyond-DFT 
approaches—are likely necessary to achieve a more comprehensive description. Future developments incorporating 
many-body techniques, such as DMFT or GW-based corrections, hold promise for refining our understanding of 
magnetic phase stability and correlation-driven phenomena in these complex intermetallics.

\begin{figure*}
\begin{center}
\includegraphics[width=\textwidth]{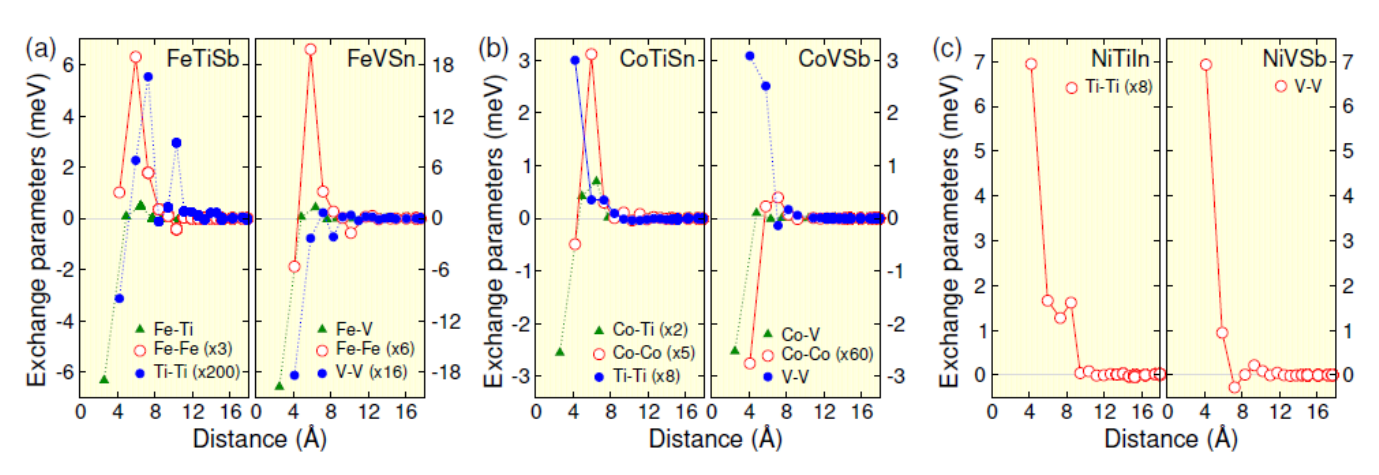}
\end{center}
\vspace*{-0.5cm} 
\caption{Intra- and inter-sublattice Heisenberg exchange parameters as a function of interatomic distance for six representative spin gapped metallic half-Heusler compounds: (a) FeTiSb and FeVSn, (b) CoTiSn and CoVSb, and (c) NiTiSn and NiVSb. Each curve corresponds to exchange interactions between specific magnetic atoms. In several cases, the exchange constants have been multiplied by an integer factor (indicated in parentheses in the legend) to enhance visibility.} 
\label{fig4}
\end{figure*}

\subsection{Magnetic interactions, spin-wave properties, and Curie temperatures} \label{sec4d}

Following the analysis of magnetic moments and Stoner instabilities presented in the previous 
sections, we now turn to the computation of magnetic exchange interactions, Curie temperatures, and 
spin-wave spectra to complete the theoretical picture of magnetism in spin gapped half-Heusler compounds. 
These quantities are essential for assessing the stability of magnetic order and the nature of 
low-energy excitations, both of which are critical for potential spintronic applications.
Our approach follows the standard methodology of mapping \textit{ab initio} total-energy calculations 
onto a classical Heisenberg Hamiltonian \cite{sandratskii1998noncollinear,nordstrom1996noncollinear,halilov1998adiabatic,van1999first,grotheer2001fast}. This approximation is well justified for systems with localized 
magnetic moments and can still yield meaningful insights for moderately itinerant magnets, provided the 
magnetic moments are not too small. However, in strongly itinerant systems, the accuracy of the Heisenberg 
mapping becomes limited. In addition, the calculation of spin-wave dispersions captures only collective 
excitations and neglects so-called Stoner excitations — single-particle spin-flip processes that can 
lead to Landau damping. This simplification is reasonable in half-metallic systems, where spin flips 
are energetically suppressed due to the position of the Fermi level relative to the minority-spin 
conduction band edge. Finally, Curie temperatures are estimated using the mean-field approximation (MFA) 
applied to the computed exchange parameters. While MFA provides qualitative trends, it is known to 
systematically overestimate the actual transition temperatures by neglecting collective spin fluctuations. 
More accurate estimates would require approaches such as the random phase approximation (RPA), which include 
spin-wave renormalization effects.

With these considerations in mind, we now proceed to the evaluation of exchange constants, which form the basis 
for our analysis of magnetic excitations and thermal stability. As described in Sec.~\ref{sec3}, the exchange 
interactions are computed using the real-space LKAG formalism, which maps the results of \textit{ab initio} electronic structure calculations onto a classical Heisenberg Hamiltonian in a linear-response framework. In magnetic 
systems such as half-Heusler compounds, two primary mechanisms contribute to the exchange interactions: 
direct and indirect exchange~\cite{Sasioglu2005}.  The direct exchange arises from interactions between magnetic 
atoms on different sublattices—typically when both transition metal atoms are magnetic and occupy nearest-neighbor positions. These intersublattice interactions are especially relevant in compounds with multiple magnetic species. 
The indirect exchange, by contrast, takes place between magnetic atoms of the same type within a single sublattice. 
It is mediated by conduction electrons and is often described by Ruderman–Kittel–Kasuya–Yosida (RKKY)–like 
mechanisms. In systems where only one sublattice hosts magnetic moments, the stability of the magnetic phase is 
governed primarily by these intrasublattice interactions.  The relative strength and sign of the direct and 
indirect exchange contributions determine the nature of the magnetic ground state and are key factors in 
establishing the Curie temperature and spin-wave behavior of the system.

Figure~\ref{fig4} presents the calculated exchange constants as a function of interatomic distance for 
six representative half-Heusler compounds. For FeTiSb and FeVSn, both with 17 valence electrons per 
formula unit, the dominant magnetic interaction is the direct intersublattice exchange between Fe and 
Ti (or V) atoms. This coupling is negative for nearest neighbors, favoring antiparallel alignment — 
consistent with the spin magnetic moments reported in Table~\ref{table1}, where Fe and Ti(V) moments 
are oriented oppositely. In contrast, the intrasublattice interactions (Fe–Fe and Ti–Ti or V–V) are 
significantly weaker; to enhance their visibility in the plot, we scaled them by factors of 3 and 
6 for Fe–Fe, and 200 and 16 for Ti–Ti and V–V, respectively.

The compounds CoTiSn and CoVSb illustrate contrasting magnetic behavior. CoTiSn exhibits itinerant 
magnetism, with magnetism primarily driven by Co–Ti exchange. The Ti–Ti interactions are much weaker, 
indicating limited magnetic contribution from the Ti sublattice. CoVSb, on the other hand, shows more 
localized magnetism, with V–V interactions that are comparable in strength to the Co–V exchange. This 
reflects a cooperative magnetic ordering involving both sublattices and is supported by the data in 
Table~\ref{table1}, where V atoms carry magnetic moments nearly an order of magnitude larger than 
those on Co.

The final two compounds, NiTiSn and NiVSb, demonstrate magnetism localized predominantly on the Ti 
and V sublattices, respectively. In NiVSb, the large spin magnetic moments on V atoms (approximately 
2\,$\mu_\mathrm{B}$) are accompanied by strong V–V exchange constants, indicating robust local-moment 
magnetism. In contrast, NiTiSn displays smaller and more delocalized Ti moments, resulting in significantly 
weaker Ti–Ti exchange. These findings highlight the diverse nature of magnetic exchange interactions 
across the half-Heusler family, governed by both electronic structure and sublattice contributions.

\begin{figure*}
\begin{center}
\includegraphics[scale=0.275]{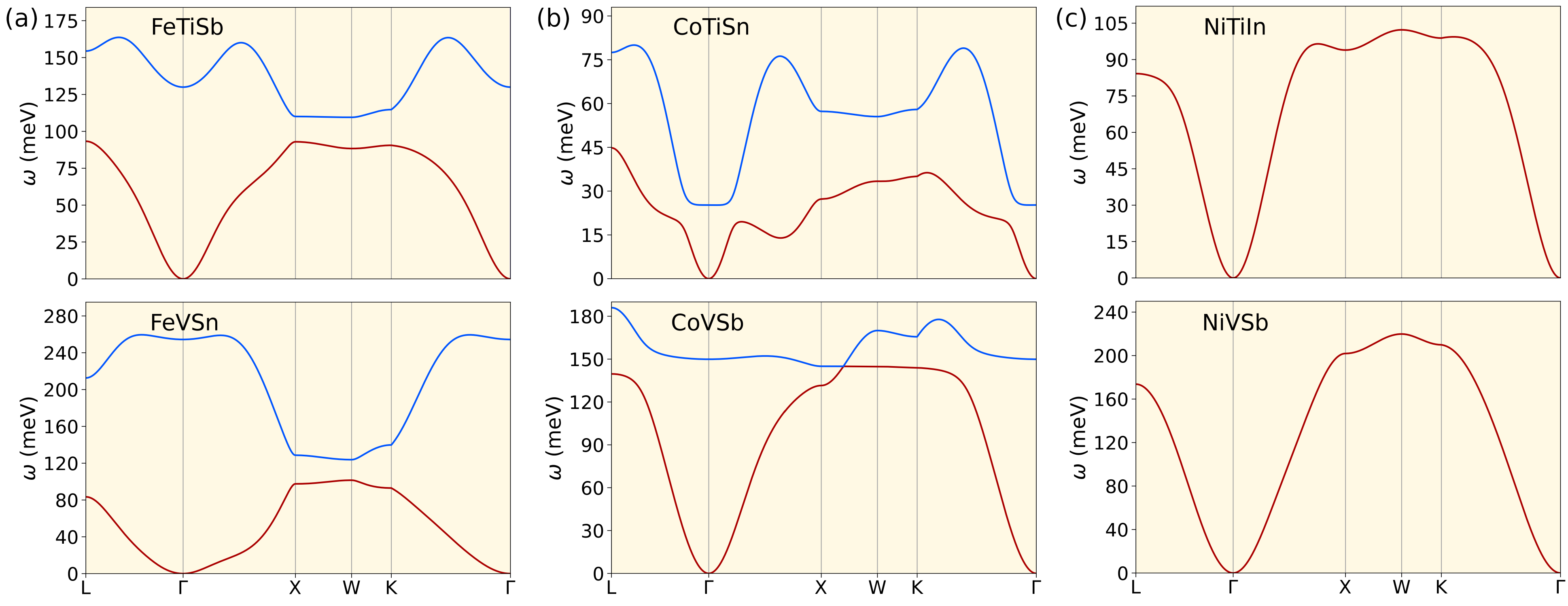}
\end{center}
\vspace*{-0.53cm} 
\caption{Calculated magnon dispersion curves for six representative half-Heusler compounds: (a) FeTiSb and FeVSn, (b) CoTiSn and CoVSb, and (c) NiTiSn and NiVSb. In compounds with only one magnetic atom per formula unit (NiTiSn and NiVSb) the spectra consist of a single acoustic magnon branch (red curve). In contrast, the other four compounds contain two magnetic atoms per unit cell, resulting in both an acoustic branch (red curve) and a higher-energy optical magnon branch (blue curve). Among them, CoTiSn exhibits ferromagnetic ordering, while FeTiSb, FeVSb, and CoVSb are ferrimagnets with antiparallel sublattice moments. All magnon spectra are plotted along the high-symmetry path L-$\Gamma$-X-W-K-$\Gamma$ in the Brillouin zone, with energy given in meV.}
\label{fig5}
\end{figure*}

The exchange constants discussed above were subsequently used to estimate the Curie temperatures, as 
outlined in Sec.~\ref{sec3}. The resulting $T_\mathrm{C}$ values are presented in the last column of 
Table~\ref{table1}. For most compounds, the Curie temperatures lie below room temperature. This is 
especially true for itinerant magnets, where $T_\mathrm{C}$ values often fall below 100\,K, reflecting 
the weaker exchange interactions associated with delocalized magnetic moments. In contrast, several 
compounds exhibiting localized magnetism show significantly higher Curie temperatures, making them 
promising candidates for spintronic applications. Notably, FeVSn, CoVSb, and NiVSb exhibit $T_\mathrm{C}$ 
values of 532\,K, 419\,K, and 684\,K, respectively.

This trend can be qualitatively understood in terms of an empirical rule suggesting that the Curie 
temperature scales with the sum of the absolute values of the atomic spin magnetic 
moments~\cite{Rusz2006,Kubler2007}. For example, in FeVSn, although the total spin moment is only 
$-0.88\,\mu_\mathrm{B}$ due to antiparallel alignment between Fe and V, the sum of the absolute moments 
is 2.88\,$\mu_\mathrm{B}$. Similarly, the corresponding values are 1.74\,$\mu_\mathrm{B}$ for CoVSb 
and 2.15\,$\mu_\mathrm{B}$ for NiVSb. While this empirical trend aligns with the relatively high 
$T_\mathrm{C}$ values in these cases, it should be emphasized that there is no strict linear 
relationship, and the applicability of this rule remains limited.

Finally, we briefly comment on the expected electronic behavior of spin gapped metals above $T_\mathrm{C}$.
In conventional magnetic metals, the loss of magnetic order typically results in a paramagnetic metallic 
state. In contrast, the spin gapped systems studied here may retain a gapped character even in the non-magnetic 
state, depending on the symmetry of the spin-resolved band structure. This scenario is particularly likely 
when both spin channels exhibit the same carrier type (either $p$-type or $n$-type), as shown in panels (d) 
and (e) of Fig.~\ref{fig1}, or when one spin channel remains semiconducting. In such cases, the paramagnetic 
state may closely resemble a gapped metal rather than a conventional one. On the other hand, spin gapped 
metals with mixed $p$- and $n$-type character [Fig.~\ref{fig1}(f)] are expected to evolve into ordinary 
paramagnetic metals above $T_\mathrm{C}$.

In addition to determining Curie temperatures, the exchange constants also allow us to investigate the 
dynamical magnetic excitations of the system—specifically, the magnon spectra, as discussed in 
Sec.~\ref{sec3}. Magnons, or spin-wave excitations, represent the low-energy collective modes of the 
spin system and are distinct from single-particle spin-flip excitations known as Stoner excitations. 
Figure~\ref{fig5} shows the calculated magnon spectra for six representative compounds, illustrating 
how the number of magnetic sublattices and the nature of magnetic ordering influence the spin-wave 
dispersion.

NiTiSn and NiVSb each have only one magnetic sublattice (Ti and V, respectively), resulting in a single 
acoustic magnon branch. This is true despite their differing magnetic character — NiTiSn being an 
itinerant magnet and NiVSb a localized one. In contrast, FeTiSb, FeVSb, CoTiSn, and CoVSb feature 
two magnetic sublattices, leading to both acoustic and optical magnon branches. The acoustic mode 
corresponds to in-phase precession of the sublattice moments, while the optical mode involves 
out-of-phase precession and appears at higher energy. Among these compounds, only CoTiSn exhibits 
ferromagnetic alignment with parallel sublattice moments, whereas the others show ferrimagnetic 
coupling with antiparallel alignment. This distinction influences both the energy separation and 
intensity distribution between the two magnon branches.

Despite differences in spectral features, the acoustic mode consistently vanishes at the $\Gamma$ 
point for all systems, as expected from the Goldstone theorem in the absence of spin-orbit coupling. 
Near the zone center, the dispersion follows a quadratic form $E(\mathbf{q}) = D \cdot |\mathbf{q}|^2$, 
where $D$ is the spin-wave stiffness constant. The calculated $D$ values, listed in Table~\ref{table1}, 
span a wide range from 165 to 936 meV\AA$^2$, reflecting the diversity of exchange interactions and 
the localized versus itinerant nature of magnetism across the studied half-Heusler compounds.

While linear spin-wave theory based on DFT-calculated exchange parameters yields valuable insights into 
collective magnetic excitations, it has inherent limitations when applied to metallic systems. In particular, 
the classical Heisenberg model employed here neglects the coupling  between collective magnons and single-particle 
spin-flip (Stoner) excitations, which can result in significant damping of spin waves. This limitation is 
especially relevant for spin gapped metallic systems, where the Fermi level may lie close to the onset of 
the spin-flip continuum. As a result, our current method does not account for Landau damping effects or 
finite magnon lifetimes that arise from such interactions. A more rigorous treatment would require going beyond 
the Heisenberg model and employing approaches such as time-dependent density functional theory (TDDFT) or 
many-body perturbation theory (MBPT), which explicitly account for dynamic electron-hole correlations \cite{savrasov1998linear,buczek2009energies,csacsiouglu2010wannier,karlsson2000spin,friedrich2014spin,costa2004theory,lounis2010dynamical}.

These limitations are particularly pertinent for compounds in which one or both spin channels exhibit 
$p$-type spin gapped metallic behavior, such as FeVSb, where the Fermi level lies close to the minority-spin 
conduction band edge. In such cases, efficient magnon–Stoner coupling can lead to pronounced damping and 
short magnon lifetimes. In contrast, compounds like CoVSb and NiVSb, where one spin channel displays 
$n$-type spin gapped metallic behavior and the other is semiconducting, are expected to exhibit more 
coherent and long-lived magnons due to the presence of a Stoner gap that separates collective modes from 
the spin-flip continuum. Similar behavior may also arise in Fe- and Ti-based half-Heuslers with a full 
semiconducting gap in one spin channel, as indicated in Table~\ref{table1}. Understanding how the type 
and size of the spin gap affect magnon coherence and damping remains an important open question, with 
direct implications for the design of next-generation spintronic and magnonic devices based on spin 
gapped metallic half-Heuslers.

\section{Summary and Conclusions}\label{sec5}

In this work, we presented a comprehensive first-principles investigation of spin gapped metallic
half-Heusler compounds, with the goal of uncovering the microscopic origin of magnetism and its 
implications for spintronic applications. Using multiple \textit{ab initio} electronic structure 
methods, we systematically explored the nature of magnetic moments, the role of electronic correlations, 
and the resulting exchange interactions across a broad family of compounds with 16, 17, 19, or 20 
valence electrons per formula unit. Our analysis of ferro-, ferri-, and antiferromagnetic configurations 
revealed that Co- and Ni-based systems predominantly exhibit itinerant magnetism, while Fe-, Ti-, 
and V-based compounds display a richer spectrum, ranging from localized to itinerant behavior—and 
in some cases, a coexistence of both. Spin-density isosurface plots further corroborate these 
trends, distinguishing localized from delocalized spin polarization in real space.

To elucidate the origin of magnetism, we estimated the Stoner parameter $I$ from cRPA-calculated 
Hubbard $U$ and Hund’s exchange $J$ values. This enabled a material-specific application of the Stoner 
criterion, which successfully accounts for the emergence of magnetism in many itinerant systems, as 
well as the absence of magnetic order in several Co- and Ni-based compounds. In the latter case, 
strong hybridization and a low DOS at the Fermi level suppress the Stoner instability, 
despite the presence of transition metal elements.

We computed Heisenberg exchange parameters using the LKAG formalism, which allowed us to 
estimate Curie temperatures and analyze spin-wave spectra. Depending on the magnetic sublattice 
configuration, magnetic order is stabilized either through inter-sublattice (direct) or intra-sublattice 
(indirect) exchange interactions. In compounds where both sublattices carry substantial spin moments, 
the Curie temperatures exceed room temperature, identifying them as promising candidates for high-temperature 
spintronic applications. The calculated magnon spectra exhibit well-defined acoustic modes and, in 
multi-sublattice systems, additional optical branches. The spin-wave stiffness constants span a 
wide range, reflecting the diversity of magnetic behavior across the half-Heusler family.

Overall, our study offers fundamental insights into the delicate interplay between localized and itinerant 
magnetism in spin-gapped metallic half-Heusler compounds. By integrating advanced electronic structure 
methods with analytical modeling, we construct a comprehensive theoretical framework for understanding 
the microscopic origins of magnetic order in these systems. This framework paves the way for the predictive
design of spintronic materials with precisely engineered magnetic and transport properties.
In particular, the coexistence of spin-resolved energy gaps and tunable magnetism uniquely positions 
spin gapped metals as promising platforms for next-generation electronic applications, including 
multifunctional steep-slope field-effect transistors and other beyond-CMOS devices.

\begin{acknowledgments}
This work was supported by several funding sources, including SFB CRC/TRR 227 and SFB 1238 (Project C01) 
of the Deutsche Forschungsgemeinschaft (DFG), the European Union (EFRE) through Grant No: ZS/2016/06/79307, 
and the Federal Ministry of Education and Research of Germany 
(BMBF) within the framework of the 
Palestinian-German Science Bridge (BMBF grant number DBP01436). 
M.T. acknowledges the TUBITAK ULAKBIM, 
High Performance and Grid Computing Center (TRUBA resources). S.G. 
acknowledges National Supercomputing Mission (NSM) for providing 
computing resources of 'PARAM SEVA' at IIT, Hyderabad, which
is implemented by C-DAC and supported by the Ministry of 
Electronics and Information Technology (MeitY) and Department of 
Science and Technology (DST), Government of India.
B.S. acknowledges financial support from Swedish Research Council (grant no. 2022-04309) and STINT Mobility Grant for 
Internationalization (grant no. MG2022-9386). The computations were enabled by 
resources provided by the National Academic Infrastructure for Supercomputing in Sweden (NAISS) at NSC 
and PDC (NAISS 2024/3-40) partially funded by the Swedish Research Council through grant agreement No. 2022-06725.

\end{acknowledgments}

\section*{Data Availability Statement}

Data available on request from the authors

\nocite{*}
\providecommand{\noopsort}[1]{}\providecommand{\singleletter}[1]{#1}%

\end{document}